\documentclass[11pt]{article}

\usepackage[final]{acl}

\usepackage{times}
\usepackage{latexsym}

\usepackage[T1]{fontenc}

\usepackage[utf8]{inputenc}

\usepackage{microtype}

\usepackage{inconsolata}

\usepackage{graphicx}

\usepackage{hyperref}
\usepackage{url}

\usepackage{booktabs}
\usepackage{algorithm}
\usepackage{algorithmic}
\usepackage{wrapfig}
\usepackage{enumitem}

\usepackage{makecell}
\usepackage{graphicx}
\usepackage{subfig}
\usepackage{multirow}
\usepackage{amsfonts}
\usepackage{amsmath}

\newcommand{\shuffle}[1]{\ensuremath{\mathsf{sf}(#1)}}
\newcommand{\indexshuffle}[2]{\ensuremath{\mathsf{sf}(#1, \pi^{#2})}}
\newcommand{\indexshufflenew}[2]{\ensuremath{\mathsf{sf}(#1, \pi_{#2})}}
\newcommand{\indexshufflenewnew}[3]{\ensuremath{\mathsf{sf}(#1, \pi^{#2}_{#3})}}
\newcommand{\ashare}[1]{[ {#1} ]}

\newcommand{\oldpi}{FME}

\newcommand{\newpi}{LOE}
\renewcommand{\paragraph}[1]{\noindent\textit{\textbf{#1.}~}}

\title{On the (In-)Security of the Shuffling Defense in the Transformer Secure Inference}

\author{
\textbf{Zhengyi Li\textsuperscript{1}},
\textbf{Yakai Wang\textsuperscript{1}},
\textbf{Kang Yang\textsuperscript{3}},
\\
\textbf{Yu Yu\textsuperscript{1,2}},
\textbf{Jiaping Gui\textsuperscript{1}},
\textbf{Yu Feng\textsuperscript{1}},
\textbf{Ning Liu\textsuperscript{1}},
\textbf{Minyi Guo\textsuperscript{1,2}},
\textbf{Jingwen Leng\textsuperscript{1,2}}
\\
\textsuperscript{1}Shanghai Jiao Tong University,
\textsuperscript{2}Shanghai Qizhi Institute,
\\
\textsuperscript{3}State Key Laboratory of Cryptology
\\
\small{
  \textbf{Correspondence:}
  \href{mailto:jgui@sjtu.edu.cn}{jgui@sjtu.edu.cn},
  \href{mailto:yangk@sklc.org}{yangk@sklc.org},
  \href{mailto:ningliu@sjtu.edu.cn}{ningliu@sjtu.edu.cn}
}
}

\begin{document}
\maketitle
\begin{abstract}
For Transformer models, cryptographically secure inference ensures that the client learns only the final output, while the server learns nothing about the client’s input. However, securely computing nonlinear layers remains a major efficiency bottleneck due to the substantial communication rounds and data transmission required. To address this issue, prior works reveal intermediate activations to the client, allowing nonlinear operations to be computed in plaintext. Although this approach significantly improves efficiency, exposing activations enables adversaries to extract model weights. To mitigate this risk, existing works employ a shuffling defense that reveals only randomly permuted activations to the client. In this work, we show that the shuffling defense is not as robust as previously claimed. We propose an attack that aligns differently shuffled activations to a common permutation and subsequently exploits them to extract model weights. Experiments on Pythia-70m and GPT-2 demonstrate that the proposed attack can align shuffled activations with mean squared errors ranging from $10^{-9}$ to $10^{-6}$. With a query cost of approximately \$1, the adversary can recover model weights with L1-norm differences ranging from $10^{-4}$ to $10^{-2}$ compared to the oracle weights.
\end{abstract}

\section{Introduction}
\label{sec:introduction}
The rapid advancement of the Generative Pre-trained Transformer (GPT)~\cite{cohen2020opengpt} has led to various applications in the real world.
Machine Learning as a Service (MLaaS) is a primary way for clients to access intelligent applications.
In MLaaS, a server hosting a private Transformer model provides a public API, allowing the client to submit data and receive inference results.
This raises privacy concerns for both clients and servers, as each party is unwilling to share their data with the other.

Secure inference using multi-party computation (MPC) or homomorphic encryption (HE) offers a privacy-preserving solution for both parties. Full-model encryption (FME) inference operates entirely on encrypted data, ensuring that the client only learns the inference result while the server learns nothing.
However, employing cryptographic protocols incurs substantial costs~\cite{cheetah,knott2021crypten}.
One major bottleneck arises from nonlinear layers. Secure computation of nonlinear layers typically utilizes piecewise polynomial approximations or iterative methods, which require multiple calls for secure multiplication, truncation, and comparison.
These operations involve tens of communication rounds and large data transmission.
Compared to linear layers, which require only a single multiplication and truncation, nonlinear layers contribute most to the overall latency.
For example, prior studies report that the secure computation of nonlinear layers accounts for $75\% \sim 90\%$ of the total latency~\cite{cho2022selective,li2023mpcformer,pang2023bolt}.

To mitigate the bottleneck on nonlinear layers, some works reveal intermediate activations to the client to evaluate nonlinear layers in plaintext. This approach is referred to as linear-only encryption (LOE) inference.
Even for complex nonlinear layers such as $\mathsf{softmax}$, $\mathsf{GELU}$, and $\mathsf{layernorm}$, their evaluation can be completed in two rounds of communication. Existing works report tens of times speedup for Transformer models, making LOE inference practically feasible. However, revealing activations to the client introduces potential leakage of model weights as a trade-off.
Early attempts to mitigate these risks, such as limiting the number of queries~\cite{GELU-Net} or using Bayesian networks~\cite{Bayhenn}, have been proven insufficient\cite{attack_revealing}.
More recent research~\cite{perm_dnn1,perm_transformer2,perm_transformer1,perm_dnn2,perm_transformer3} introduces the insight that computing nonlinear layers is independent of the positions of values in the activation tensor. This insight enables a more robust defense by revealing shuffled intermediate activations to the client. When permuting a tensor with $d$ elements, the probability that a client correctly guesses the permutation is $1/(d!)$, which is negligibly small for Transformer models with hundreds of dimensions. Consequently, shuffled activations are considered difficult to exploit by adversaries to exploit in existing black-box attacks~\cite{carlini2020cryptanalytic,steal_gpt}.

This work challenges the security claims of the shuffling defense applied to Transformer models.
We show that model weights can be extracted as long as the adversarial client is able to align differently shuffled activation vectors to a common permutation, even when this permutation remains unknown. To achieve this, the adversary queries the Transformer to generate activation vectors that are differently shuffled but numerically close. Since shuffling preserves the values, this proximity enables the recovery of element-wise correspondences across vectors. The aligned activations can then be used to extract model weights by solving a linear system.
As the true permutation is unknown to the adversary, the recovered weights differ from the original weights by row-wise and column-wise permutations. Despite this discrepancy, we show that the extracted weights preserve computational correctness during forward propagation.

The extracted weights are useful for several reasons. First, they closely approximate the original weights and can be used to analyze the private model. Second, they strengthen existing black-box attacks, such as by serving as effective initialization for surrogate models.
Overall, the proposed attack reveals information leakage beyond what is assumed under the shuffling defense, calling for a reevaluation of its security in secure inference and related privacy-preserving settings~\cite{shen2022soter,shuffleforDP4}.

Our contributions are summarized as follows:
\begin{itemize}[leftmargin=10pt]
    \item We demonstrate that an adversary can successfully align activation vectors that are shuffled differently but have close values. Moreover, to address the issue of discrete inputs in Transformer models that cannot generate intermediate activations with close values, we leverage the probabilistic errors inherent in the secure truncation protocol to inject the desired perturbation.
    
    \item With the aligned activations, we theoretically show that the adversary can  extract model weights that differ from the original weights up to row-wise and column-wise shuffling. 
    Despite the row and column shuffling, the extracted weights perform the same forward propagation as the original weights, even if the adversary does not know the specific permutation. 
    
    \item  We evaluate the effectiveness of the proposed attack on two Transformer architectures widely used in secure inference, Pythia-70m and GPT-2. Experimental results show that we can align the shuffled activations with mean squared errors  ranging from $10^{-9}$ to $10^{-6}$. With a total query cost of approximately one dollar, the adversary can extract model weights whose L1-norm differences from the original weights range from $10^{-4}$ to $10^{-2}$.
\end{itemize}

\section{Preliminaries}
\label{sec:background}
This work examines the secure inference of the Generative Pre-trained Transformer (GPT)~\cite{cohen2020opengpt}. There is a client with a private input and a service provider (server) with a private model. The model architecture is required to be known by both parties to interactively execute the protocols. The server aims to hide the model weights, while the client wants to protect the input.


\subsection{Full-model-Encryption Secure Inference }
\paragraph{Privacy goal of the \oldpi{}}
Secure inference based on homomorphic encryption (HE) and multi-party computation (MPC) provides a solution to protect the privacy of both servers and clients. 
Secure inference with full-model encryption (\oldpi{}) achieves the following privacy goals: the client $P_1$ only learns the inference results, while the server $P_0$ knows nothing about the client's input.

\begin{figure}[tbp] 
  \centering
    \includegraphics[width=0.99\linewidth]{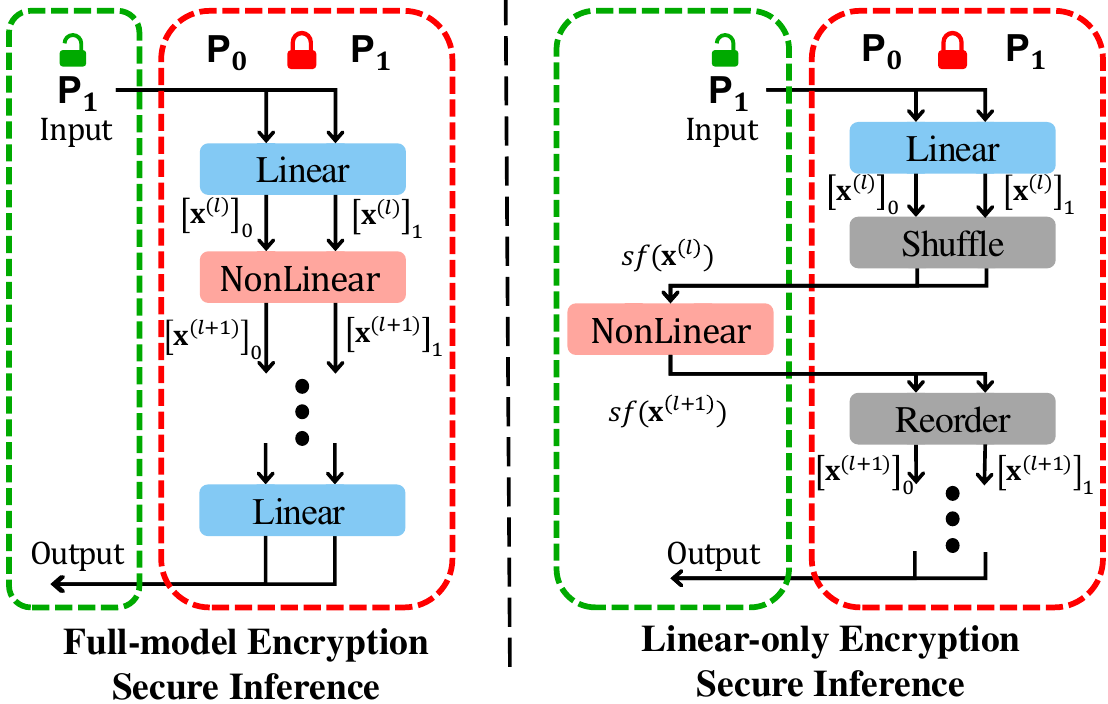}
  \caption{The red locked part is private to the owner or invisible for both parties, and the green unlocked part is revealed to the client. \oldpi{} inference only reveals the final output to the client. \newpi{} inference also reveals shuffled activations to the client.}
  \label{fig:illustrate_private_inference}
\end{figure}

\paragraph{Explain \oldpi{} Inference}
Figure \ref{fig:illustrate_private_inference} illustrates the \oldpi{} secure inference~\cite{li2024nimbus,kei2025shaft,park2025powerformer,moon2025thor} that operates entirely on encrypted inputs and model weights, and only the final output is revealed to the client.
Each layer's inputs and outputs are additive secret shares $[\cdot]_i$, allowing for flexible linkage of different layers.
The additive secret share~\cite{ito1989secret} over the ring $\mathbb{Z}_{2^{\ell}}$ is defined as follows: for a given value $x \in \mathbb{Z}_{2^{\ell}}$, the client and server each hold two random shares $\ashare{x}_0 \in \mathbb{Z}_{2^{\ell}}$ and $\ashare{x}_1 \in \mathbb{Z}_{2^{\ell}}$ such that $x = \ashare{x}_0 + \ashare{x}_1 \mod 2^{\ell}$ holds.
To convert the floating-point value $x_f$ of the model to the ring value $x$, it is scaled by a large factor and rounded to the nearest integer: $x=\left\lfloor x_f \cdot 2^{p} \right\rceil$ for $p$-bit precision. The decoding of $x$ is computed as $x_f = x / 2^{p}$.

\paragraph{Nonlinear Layers as the Bottleneck}
The primary efficiency bottleneck for cryptographic methods lies in nonlinear layers.
In contrast to linear layers, which typically require only two rounds of communication, nonlinear layers usually demand dozens of rounds of communication along with a significant amount of data transmission.
In secure computations, the costs associated with communication rounds and data transmission are the primary bottleneck.
Consequently, the considerable communication overhead makes nonlinear layers become the major bottleneck in secure inference, accounting for approximately $75\% \sim 90\%$ of the communication~\cite{cho2022selective,li2023mpcformer,pang2023bolt}, especially under suboptimal network conditions.

\subsection{Linear-only-Encryption Secure Inference}
\label{subsec:loe_inference}
\paragraph{Explain \newpi{} Inference}
Despite numerous efforts~\cite{cryptflow2,rathee2021sirnn,pang2023bolt,li2024nimbus} to enhance protocol efficiency or replace nonlinear layers with cryptography-friendly alternatives~\cite{mishra2020delphi,cho2022selective,THE-X,li2023mpcformer}, nonlinear layers remain the major bottleneck. 
Another category of works aim to find a balance between performance and privacy.
They propose linear-only encryption (\newpi{}) inference, which reveals  intermediate activations to the client, enabling the computation of nonlinear layers in plaintext~\cite{GELU-Net,Bayhenn,perm_dnn1,perm_dnn2,perm_transformer3}. This approach allows nonlinear layers to require only two rounds of communication, regardless of their complexity.
\newpi{} inference shows tens of times speedup compared to \oldpi{} inference, effectively advancing secure inference to a practical solution.

\paragraph{Shuffling defense} 
As a trade-off for performance, revealing activations to the client may expose model weights.
In a typical deep neural network model, which consists of interleaved linear and nonlinear layers, revealing activations makes both the input and output of the linear layers accessible to the client. Consequently, the weights can be extracted by solving a linear system.
Early attempts to mitigate this risk, such as limiting the number of queries~\cite{GELU-Net} or using Bayesian networks~\cite{Bayhenn}, have proven insufficient\cite{attack_revealing}.

Later works propose the shuffling defense~\cite{perm_dnn1,perm_dnn2,perm_transformer1,perm_transformer3} to enhance the performance-privacy trade-off, as illustrated in Figure \ref{fig:illustrate_private_inference}.
They provide insight that correctly computing nonlinear layers is independent of the positions of values in the activation tensor. Some nonlinear layers, such as $\mathsf{ReLU}$ and $\mathsf{GELU}$, support shuffling across the entire activation tensor. Other nonlinear layers, like $\mathsf{softmax}$ and $\mathsf{LayerNorm}$, support only row-wise shuffling due to information reduction along the last dimension.
Based on this insight, the activations are randomly shuffled before being revealed to the client.
The results of nonlinear layers are then re-shared and reordered among the parties. The shuffling defense does not introduce extra communication overhead since shuffling and reordering can be performed locally.

\paragraph{Prior security analysis}
Prior works claim~\cite{perm_dnn1,perm_dnn2,perm_transformer1} recovering the shuffling is practically impossible when the number of elements is not too small. 
For a vector with hidden dimension $h$, there are $h!$ possible permutations. In Transformer models, where $h \geq 100$, there are at least $100! \approx 10^{157}$ possible permutations. Thus, the probability of correctly guessing the permutation is negligible. 
These works also show original activations and shuffled activations have minimal correlation, and thus existing black-box attacks fail to exploit the destroyed information for meaningful attack.
Note prior works do not regard permutation among different vectors as a defense. For example, the intermediate activation in the GPT decoding only contains one vector. In other Transformer applications, the adversary can also bypass the permutation in this dimension by feeding a single token as input.


\subsection{Transformer Model}
This work focuses on the Transformer~\cite{transformer} models, especially the widely used GPT model~\cite{cohen2020opengpt}.
The main body of Transformer models comprises stacked Transformer encoders/decoders, each containing an attention module and a feed-forward network (FFN) module. 
Existing \newpi{} works~\cite{perm_transformer1,perm_transformer2,perm_transformer3} allow the $\mathsf{softmax}$, $\mathsf{GELU}$, and $\mathsf{layernorm}$ layers to be evaluated in plaintext.


\paragraph{Attention Module}
The attention module starts with a linear layer ${\sf Linear}_{qkv}$, which projects the input into three independent activation tensors: $\mathbf{Q}$, $\mathbf{K}$, and $\mathbf{V}$. 
The multi-head attention mechanism splits them into $H$ heads and parallelly compute all heads. The $h_{th}$ head is computed as:
\begin{equation} 
\resizebox{\columnwidth}{!}{$
\mathbf{Attn}(\mathbf{Q}^h, \mathbf{K}^h, \mathbf{V}^h)=\mathsf{softmax}\left(\frac{\mathbf{Q}^h  {\mathbf{K}^h}^{T}}{\sqrt{d_k}}\right)  \mathbf{V}^h,
$}
\label{equ:attention}
\end{equation}
where $d_k$ is the hidden dimension of the $\mathbf{K}^h$. 
The outputs of all heads are concatenated and fed into another linear layer $\operatorname{\sf Linear}_o$. Finally, a residual connection and a normalization layer are applied to generate the module's output. The $\mathbf{K}$ and $\mathbf{V}$ of prefix tokens are cached as the KV cache to save the redundant computation~\cite{kv_cache}.

\paragraph{FFN Module} The FFN module consists of two linear layers and one activation layer, structured as: 
\begin{equation}
\mathbf{FFN}(\mathbf{X})=\operatorname{\sf Linear}_{h_2}(\operatorname{\sf ACT}(\operatorname{\sf Linear}_{h_1}(\mathbf{X}))),
\end{equation}
where $\mathsf{ACT}$ is the activation function, such as $\mathsf{GELU}$ and $\mathsf{ReLU}$. 
Similar to the attention module, the output incorporates a residual connection and a normalization layer.

        
        

\section{Problem Formulation}
\label{sec:problem_formulation}
\paragraph{Notations}
We use bold lowercase letters to represent vectors, such as the input of the $l_{th}$ linear layer $\mathbf{x}^{(l)}$, and the output probability distribution $\mathbf{y}$. The bold uppercase letters represent the matrix, such as the weights of the $l_{th}$ linear layer $\mathbf{W}^{(l)}$.
The permutation matrix is a square matrix obtained by permuting the rows (or columns) of the identity matrix~\cite{permutation_matrix}.
Shuffling columns of a vector or matrix is represented by multiplying a permutation matrix $\pi$ on the right, denoted as $\indexshuffle{\mathbf{x}}{} = \mathbf{x} \pi$.
For simplicity, we denote it as $\shuffle{\mathbf{x}}$ when the specific permutation is not important in the context.

\paragraph{MPC Setup}
We follow the semi-honest threat model, which is widely used in both \oldpi{} and \newpi{} secure inference~\cite{mohassel2018aby3,perm_dnn1,perm_dnn2,pang2023bolt,hao2022iron}. In this threat model, the client and server are curious about each other's information but strictly adhere to cryptographic protocols. The shuffling defense in \newpi{} inference works in either a 2-party or 3-party setting and can be generalized to different cryptographic protocols. The proposed attack is also applicable to different numbers of parties and cryptographic protocols. Later discussion focuses on the 2-party setting and protocols commonly used in prior \newpi{} works as an example~\cite{perm_dnn2,perm_transformer1}.

\begin{figure}[tbp] 
  \centering
    \includegraphics[width=0.8\linewidth]{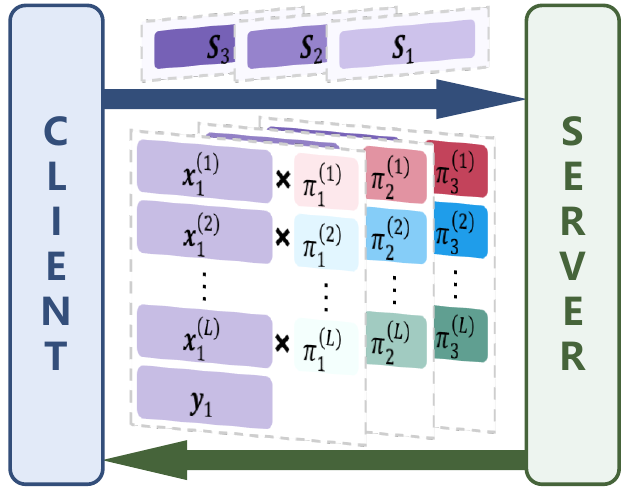}
  \caption{The interface of \newpi{} inference.}
  \label{fig:threat_model}
\end{figure}

\paragraph{Threat Model}
We use the GPT to illustrate our attack. We define the \newpi{} inference's interface to the model by $\mathcal{O}$, as illustrated in Figure~\ref{fig:threat_model}. It functions as $\left[ \indexshufflenewnew{\mathbf{x}^{(1)}_i}{(1)}{i}, \indexshufflenewnew{\mathbf{x}^{(2)}_i}{(2)}{i}, \cdots, \mathbf{y}_i \right] = \mathcal{O} (\mathbf{S}_i)$, where $i$ is the query index. The input $\mathbf{S}$ is a prefix that includes the client prompt and previously generated tokens. The client obtains the shuffled activation vector $\indexshufflenewnew{\mathbf{x}^{(l)}_i}{(l)}{i}$ of all linear layers, along with the final probability $\mathbf{y}_i$ of the generated token. 
\textit{Note that for activation vectors at the same layer depth, the applied permutations are random for different model inputs.}
Other requirements follow those commonly used in existing model attacks, such as allowing free choice of input and permitting a mild number of queries.

\section{Extraction Attack of the \newpi{} Inference}
\label{sec:attack}

Section \ref{subsec:break_shuffle} introduces the proposed attack that aligns distinct shuffled activation vectors without knowing the actual permutation. Furthermore, Section \ref{subsec:gpt_alignment} addresses the unique challenges associated with alignment in Transformer models. Finally, Section \ref{subsec:expose_weights} demonstrates that the aligned vectors enable the extraction of model weights that perform the same functionality as the original model. The complete attacking algorithm is in Appendix \ref{appendix:complete_attack}.

\subsection{Breaking shuffling by Vector Alignment}
\label{subsec:break_shuffle}
\paragraph{Alignment Definition}
Suppose the adversarial client holds two distinct shuffled activation vectors $\mathbf{x}_a^{\prime}=\indexshufflenew{{\mathbf{x}}_{a}}{a}$ and $\mathbf{x}_b^{\prime}=\indexshufflenew{{\mathbf{x}}_{b}}{b}$.
The objective of the alignment is to rearrange the elements in vectors such that the corresponding elements are matched up, even though the adversary remains unaware of specific permutations $\pi_a$ and $\pi_b$.
The alignment can be mathematically formalized as transforming the permutation $\pi_b$ of $\mathbf{x}_b^{\prime}$ to  $\pi_a$ of $\mathbf{x}_a^{\prime}$ through a transformation matrix $\mathbf{M}$, which is also a permutation matrix~\cite{permutation_matrix}, such that $\pi_a = \pi_b*\mathbf{M}$ or $\indexshufflenew{{\mathbf{x}}_{b}}{a}=\indexshufflenew{{\mathbf{x}}_{b}}{b}*\mathbf{M}$.

\begin{figure}[tbp] 
  \centering
    \includegraphics[width=0.99\linewidth]{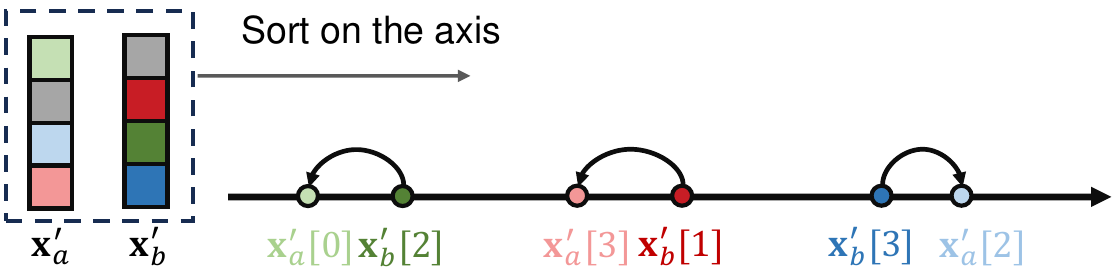}
  \caption{Aligning two shuffled vectors based on the closest distance. Three out of four pairs of elements are drawn to illustrate their correspondence for two sufficiently close vectors.
  }
  \label{fig:matching_attack}
\end{figure}
\paragraph{Aligning Close Activation Vectors}
Directly guessing the transformation matrix $\mathbf{M}$ of two unrelated activation vectors requires enumerating all possible permutation matrices. This is impractical and the adversary cannot verify whether a transformation matrix is correct. 
In contrast, the proposed attack considers the alignment of vectors $\mathbf{x}_a$ and $\mathbf{x}_b$ that are $\epsilon$-close, such that $\|\mathbf{x}_a$ - $\mathbf{x}_b\| < \epsilon$.
With the closely aligned activation vectors, the proposed attack solves $\mathbf{M}$ by leveraging the proximity between corresponding elements, as illustrated in Figure \ref{fig:matching_attack}.
The elements of the two shuffled vectors are positioned along the axis, with similar colors indicating correspondences between elements.
Since shuffling preserves the original values, when the two original vectors are sufficiently close, the distance between corresponding elements in the shuffled vectors remains much smaller than the distance between non-corresponding elements.
With this property, solving the transformation matrix $\mathbf{M}$ is equivalent to minimizing the following objective\begin{equation}
    \min_{\mathbf{M}} \| \mathbf{x}_a^{\prime}-\mathbf{x}_b^{\prime} * \mathbf{M} \|_2.
\label{equ:transforming_pi}
\end{equation}
An extreme case is when the two $\epsilon$-close model inputs have $\epsilon=0$, enabling trivial alignment by matching elements with identical values. 
In general cases when $\epsilon>0$, solving $\mathbf{M}$ using Equation \eqref{equ:transforming_pi} is difficult as the elements in $\mathbf{M}$ are discrete. Therefore, we reformulate it as follows and defer the proof of equivalence to Appendix \ref{appendix:equivalence_proof}.
\begin{equation}
\begin{aligned}
\min_{\mathbf{M}} &  \sum_{i,j\in \mathcal{H}}\mathbf{M}[i, j] \odot \mathbf{D}[i, j] \\
\text { s.t. } & \sum_{i \in \mathcal{H}}  \mathbf{M}[i, j]=1 \text { for } j \in \mathcal{H}, \\
& \sum_{j \in \mathcal{H}} \mathbf{M}[i, j]=1 \text { for } i \in \mathcal{H}
\end{aligned}
\label{equ:matching}
\end{equation}
The operator $\odot$ denotes the Hadamard product. Suppose both vectors contain  $h$ elements, and $\mathcal{H}=\{1,2, \cdots h\}$ is the set of possible values of $i$ and $j$.
$\mathbf{D} \in \mathbb{R}^{h \times h}$ is a matrix whose element $\mathbf{D}[i,j] = (\mathbf{x}_a^{\prime}[i]-\mathbf{x}_b^{\prime}[j])^2$.
The constraint is because $\mathbf{M}$ is a permutation matrix. Each element $\mathbf{M}[i, j] \in \{0,1\}$ can be interpreted as whether the $i_{th}$ element of $\mathbf{x}_b^{\prime}$ corresponds to the $j_{th}$ element of $\mathbf{x}_a^{\prime}$.
This is a standard assignment problem that can be optimally solved using the Hungarian algorithm~\cite{linearMatching}. 

\subsection{Vector Alignment in GPT}
\label{subsec:gpt_alignment}
This section details how the proposed alignment works in a real Transformer model. We describe the alignment of differently shuffled activation vectors at a certain layer depth, and the procedure is equally applicable to activations at other layer depths.

\paragraph{Constructing Close Activation vectors}
To construct the close intermediate activation vectors required by the proposed alignment, we exploit the Lipschitz continuity of neural network models~\cite{virmaux2018lipschitz}. Given two $\epsilon$-close model inputs $\mathbf{S}_1$ and $\mathbf{S}_2$, such that $\|\mathbf{S}_1$ - $\mathbf{S}_2\| < \epsilon$, the following inequality holds:
\begin{equation}
    \| \mathbf{x}_{ a }- \mathbf{x}_{ b } \| \leq \mathcal{L} \| S_ { a } - S_ { b } \| < \mathcal{L}\epsilon,
\end{equation}
where $\mathcal{L}$ is the Lipschitz constant. It bounds the difference between intermediate activations $\mathbf{x}_1$ and $\mathbf{x}_2$ based on the difference between the model inputs $\mathbf{S}_1$ and $\mathbf{S}_2$. 

In this way, the adversarial client initiates the attack by querying the Transformer model with $n$  model inputs that are $\epsilon$-close to each other.
When generating the next token, at a certain layer depth, the client acquires $n$ activation vectors $\indexshufflenew{{\mathbf{x}}_{1}}{1}, \indexshufflenew{{\mathbf{x}}_{2}}{2},\cdots, \indexshufflenew{{\mathbf{x}}_{n}}{n}$ that exhibit similar magnitudes ($\|x_i-x_j\| < \mathcal{L}\epsilon$ for any $i,j \in \{1,2,\cdots,n\}$) but distinct permutations. The client then selects an arbitrary $k \in \{1,\cdots,n\}$ and uses permutation $\pi_{k}$ of $\indexshufflenew{{\mathbf{x}}_{k}}{k}$ as the reference. Aligning the remaining vectors to the chosen vector can be represented as $ \indexshufflenew{{\mathbf{x}}_{1}}{1}, \indexshufflenew{{\mathbf{x}}_{2}}{2},\cdots, \indexshufflenew{{\mathbf{x}}_{n}}{n}    \xrightarrow{Alignment} \indexshufflenew{{\mathbf{x}}_{1}}{k}, \indexshufflenew{{\mathbf{x}}_{2}}{k},\cdots, \indexshufflenew{{\mathbf{x}}_{n}}{k}$.
The vectors aligned with the common permutation $\pi_k$ are organized into an matrix $\mathbf{X}\pi_k$ with $n$ rows, which is later used for model weight extraction.
It is crucial to emphasize that the specific permutation $\pi_k$ remains unknown to the client. The extraction of model weights does not require explicit knowledge of $\pi_k$, which we explain in Section \ref{subsec:expose_weights}.

\paragraph{Challenge of Constructing Close Inputs in GPT} 
The alignment process in Transformer models presents unique challenges in constructing $\epsilon$-close model inputs. Transformer models accept discrete token indices as inputs, which are transformed into floating-point hidden vectors through the embedding layer. This constraint fails the adversary's ability to manipulate model inputs freely to achieve the desired $\epsilon$-close conditions. Moreover, directly manipulating intermediate activations and redistributing them among parties is infeasible, as this approach violates the semi-honest threat model assumed in existing \newpi{} frameworks, thus make the attack trivial.

To overcome this limitation, we exploit a seemingly innocuous property of secure inference to construct close intermediate activations. To execute secure inference protocols, floating-point values in Transformer models are encoded as fixed-point values on the ring, and a truncation step is necessary to reset the fixed-point precision after multiplication. To maintain model accuracy, secure inference frameworks typically employ high precision; for example, the SecretFlow-SPU framework~\cite{ma2023secretflow} defaults to 18-bit precision. We notice existing secure truncation protocols introduce a stochastic 1-bit error~\cite{mohassel2018aby3,cryptflow2,ma2023secretflow}.  Then such 1-bit error manifests as a perturbation of magnitude $2^{-18}$, which we strategically exploit to generate close intermediate activations. Consequently, as the adversary repeatedly feeds the same prompt, the truncation process inherently introduces the desired minor perturbations into the intermediate activations.

\subsection{Exposing Model Weights}
\label{subsec:expose_weights}
With the aligned activation vectors, this section explains how the adversary derives weights of linear layers. 
For the $l_{th}$ linear layer, its weights satisfy the following equation
\begin{equation}
    \mathbf{X}^{(l+1)} =\mathbf{X}^{(l)}  \mathbf{W}^{(l)},
    \label{equ:general_linear_system}
\end{equation}
where $\mathbf{W}^{(l)} \in \mathcal{R}^{h_{in} \times h_{out}}$ is the weight of  $l_{th}$ linear layer. 
The standard way to solve weight is through $\mathbf{W}^{(l)} = \mathbf{X}^{(l)^{-}} \mathbf{X}^{(l+1)}$, where $\mathbf{X}^{(l)^{-}}$ is the inverse of $\mathbf{X}^{(l)}$.  To guarantee a meaningful inverse matrix, the adversary needs at least $h_{in}$ input-output vector pairs (assuming full-rank weight matrix).

In the proposed attack, the adversary obtains inputs and outputs in the form of $\mathbf{X}^{(l)} \pi^{(l)}$ and $\mathbf{X}^{(l+1)} \pi^{(l+1)}$. Using these aligned matrices, the adversary extracts weights by 
\begin{equation}
\resizebox{\columnwidth}{!}{$
    \mathbf{W}^{{(l)}^{\prime}} = \pi^{(l)^{-}} \mathbf{X}^{(l)^{-}} \mathbf{X}^{(l+1)} \pi^{(l+1)} = \pi^{(l)^{-}} \mathbf{W}^{(l)} \pi^{(l+1)}.
    $}
    \label{equ:shuffled_general_linear_system}
\end{equation}
The appendix \ref{subsec:handle_multi_head_attention} also addresses the case when the matrix multiplication of activations in the attention module is privately computed.

\paragraph{Functionality Invariance to Shuffling}
In equation \eqref{equ:shuffled_general_linear_system}, the adversary's solved weights $\mathbf{W}^{(l)'}$ differ from the original weights $\mathbf{W}^{(l)}$ by row-wise and column-wise shuffling.
Importantly, the stolen permuted weights perform the same functionality as the original weights. This is because the weight matrices of the neural network are inherently invariant to such shuffling ~\cite{carlini2020cryptanalytic,jagielski2020high}. For solved weight matrices of consecutive layers, as long as the output dimension of the former layer and the input dimension of the latter layer follow the same permutation, the functionality of forward propagation remains correct. We provide a detailed explanation in the Appendix \ref{appendix:proof_equivalent_weights}.

\paragraph{Handling the Ill-conditioned Problem}
Although Equation \eqref{equ:shuffled_general_linear_system} is theoretically solvable, practical computation introduces additional considerations.
The proximity of the aligned activation vectors makes the computation of the inverse of the input matrix $\mathbf{X}^{(l)} \pi^{(l)}$ ill-conditioned~\cite{neumaier1998solving}. The ill-conditioning is measured by the condition number $\sigma_{max} / \sigma_{min}$, where $\sigma_{max}$ and $\sigma_{min}$ represent the maximum and minimum singular values of the input matrix. A large condition number indicates that numerical errors from the machine during computation are amplified, leading to significant inaccuracies when computing the matrix inverse. To mitigate such inaccuracies, the adversary needs to manually set an upper bound $\mathcal{C}$ on the condition number. Consequently, when computing the pseudo-inverse of the input matrix $\mathbf{X}^{(l)} \pi^{(l)}$, singular values smaller than $\sigma_{max} / \mathcal{C}$ are discarded to ensure a stable approximation.
In future work, the accuracy of the solution can be further enhanced by utilizing higher machine precision or adopting iterative methods~\cite{richardson1911ix}.

\paragraph{Query Complexity}
The attack complexity of the query number is highly efficient at $O(N)$, where $N$ is the largest rank of the weight matrix in the model, typically corresponding to the dimension of the FFN module.
The efficient linear complexity is because 1) our attack finally reduces to solving a linear system, where the minimum number of required vectors exceeds the weight matrix rank, and 2) each query simultaneously gathers one activation vector for all layer depths.


\section{Evaluation}
\label{sec:experiments}

\subsection{Experimental Setup}

\paragraph{Victim Transformer Models}
We validate the proposed attack on two Transformer models: Pythia-70m~\cite{pythia} and GPT-2~\cite{gpt2}. These model scales have been widely employed in prior works on secure Transformer inference~\cite{THE-X,perm_transformer3,pang2023bolt,li2024nimbus}.
We demonstrate the weight stealing results of all types of linear layers in the Transformer model, including $\mathbf{W}_{qkv}$, $\mathbf{W}_{o}$, $\mathbf{W}_{h1}$, and $\mathbf{W}_{h2}$. 

\paragraph{Query Number} 
Theoretically, the number of required queries to compute the input matrix inverse is the weight's dimension.
In our experiments, due to the ill-conditioned problem, we set the query number to 16 times the model's maximum dimension to improve the accuracy of the matrix inverse in numerical computation. The query count for Pythia-70m with the maximum dimension 2048 is 32768. The query count for GPT-2 with the maximum dimension 3072 is 49512. Most GPT APIs charge users on a per-token basis. Since our attack is independent of response length or content, the adversary can feed a query that receives very short responses, such as having the model respond only "yes" or "no".
Using the commercial API pricing~\cite{openai_pricing}, the estimated costs for the attack on both Pythia-70m and GPT-2 remain under one dollar, demonstrating practical feasibility.


\paragraph{Fixed-point (FXP) Precision}
In secure inference, fixed-point precision of the ring number is originally intended to ensure the fixed-point errors within the tolerance of the model. 
For such an innocuous property, our attack algorithm strategically exploits the fixed-point error to construct close activation vectors. 
Thus, the choice of fixed-point precision impacts the effectiveness of the attack algorithm. 
In the well-established secure inference framework, such as SecretFlow-SPU~\cite{ma2023secretflow}, the default precision of 18 bits is sufficient to maintain computational accuracy. To further evaluate the robustness of our attack, we examine smaller precisions, specifically 14 and 16 bits.

\subsection{Alignment Error}
\label{subsec:alignment_result}
\begin{table}[tbp]
\footnotesize
  \centering
  \caption{
  For input and output of the layer, the table lists counts of correctly corresponded elements and mean squared errors of aligned vectors compared to the oracle-aligned vectors.}

\resizebox{\columnwidth}{!}{%
\begin{tabular}{cccccc}
\toprule
(Input,Output) & Fxp   & \multicolumn{2}{c}{Input } & \multicolumn{2}{c}{Output} \\
\cmidrule{3-6}Dimension & Pre.  & Count & Error & Count & Error \\
\midrule
\multirow{3}[2]{*}{(512,2048)} & 14    & 508   & 9.4E-07 & 1996  & 1.2E-06 \\
      & 16    & 511   & 1.2E-08 & 2040  & 9.1E-08 \\
      & 18    & 512   & 0.0E+00 & 2046  & 4.0E-08 \\
\midrule
\multirow{3}[2]{*}{(2048,512)} & 14    & 2004  & 3.0E-06 & 504   & 5.9E-06 \\
      & 16    & 2038  & 3.6E-07 & 510   & 1.2E-07 \\
      & 18    & 2046  & 5.5E-08 & 511   & 1.5E-08 \\
\midrule
\multirow{3}[2]{*}{(768,3072)} & 14    & 756   & 6.0E-07 & 3052  & 1.2E-07 \\
      & 16    & 766   & 2.4E-08 & 3066  & 6.4E-09 \\
      & 18    & 766   & 2.5E-08 & 3070  & 3.7E-09 \\
\midrule
\multirow{3}[2]{*}{(3072,768)} & 14    & 3056  & 2.5E-07 & 762   & 5.8E-08 \\
      & 16    & 3068  & 1.0E-08 & 764   & 1.9E-08 \\
      & 18    & 3070  & 1.2E-08 & 764   & 5.0E-09 \\
\bottomrule
\end{tabular}%
}

  \label{tab:alignement_error}%
\end{table}%

\begin{figure*}[tbp] 
	\centering
    \includegraphics[width=0.99\linewidth]{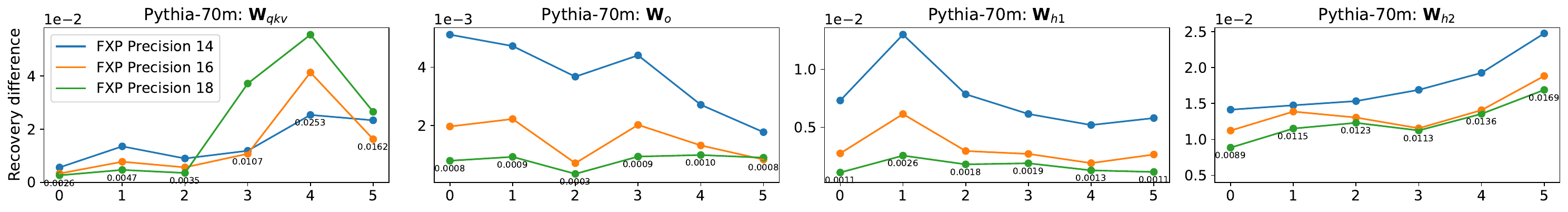}
    
    \includegraphics[width=0.99\linewidth]{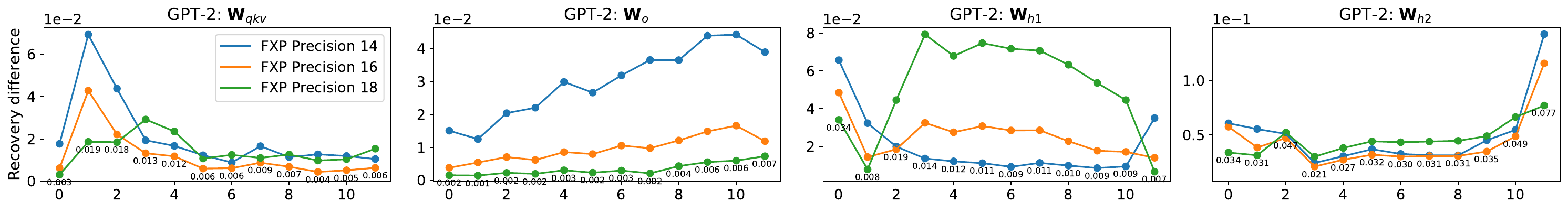}

	\caption{The attacking results on Pythia-70m and GPT-2. The x-axis indicates the depth of the decoders, and the y-axis indicates the L1-norm difference between the original and extracted weights (up to a row and column shuffling). Different colors correspond to different fixed-point precisions.
	}
	\label{fig:e2e_attack_results}
\end{figure*}

Table \ref{tab:alignement_error} presents the counts of correctly corresponded elements and mean squared errors (MSE) compared to the oracle-aligned activation vectors. 
The evaluation focuses on the $\mathsf{Linear}_{h1}$ and $\mathsf{Linear}_{h2}$ of Pythia-70m and GPT-2 at a certain depth. They are selected because their input dimensions represent the maximum and minimum cases among all linear layer types, corresponding to the highest and lowest difficulty. 
The results are averaged across the results of all obtained activation vectors.
For all precisions, less than 2\% of the elements are mismatched, with higher fixed-point precision resulting in a higher correct alignment ratio. This is due to increased precision reducing perturbations in the intermediate activations, enabling more accurate element matching. The magnitudes of MSE range from $10^{-9}$ to $10^{-6}$, primarily caused by rare mismatches and occur only when the mismatched element is close to the correct corresponding element. This accurate alignment builds a solid foundation for later extracting weights.

\subsection{Results on Extracted Model Weights}

Figure \ref{fig:e2e_attack_results} shows the L1-norm difference between the original and extracted weights. 
The extracted weights differ from the original ones by a random permutation of a row and column, hence we report the difference between each value in the extracted weights and its counterpart in the original weights.
The results indicate our attack successfully extracts weights with minor errors. This provides an accurate approximation of the original weights for malicious usage. For example, the extracted weights can be directly used to analyze model information or as an initialization for training the surrogate model. Below, we analyze the attack results in detail.
In most cases, our attack successfully extracts weights with L1-norm difference ranging from $10^{-4}$ to $10^{-2}$. We find the smaller model Pythia-70m consistently exhibits lower L1-norm differences compared to GPT-2. A common intuition regarding neural network attacks is that the difficulty of the attack increases with model size~\cite{carlini2020cryptanalytic,steal_gpt}. The same principle applies to our attack, as the difficulty to extract a weight depends on its input dimension. Similarly, the L1-norm differences for $\mathbf{W}_{h1}$ with input dimension of $d_{model}$ are generally smaller than that for $\mathbf{W}_{h2}$ with input dimension of $4*d_{model}$.

Higher precision generally results in smaller L1-norm differences. This is because higher precision leads to more accurate alignment, which benefits solving the linear system. 
However, there are cases where a lower precision of 14 results in smaller errors. This occurs because lower precision means more diverse vectors in the input matrix and relieves the ill condition of the problem,
leading to improved approximations of the matrix inverse when numerically solving the linear system.

\subsection{Singular Value Analysis}
\label{subsec:svd_analysis}
As Section \ref{subsec:expose_weights} shows, the proximity in activation matrices results in a high condition number  $\sigma_{max}/\sigma_{min}$, which means inaccuracies of matrix inverse and solved weights. 
\begin{figure}[tbp] 
  \centering
    \includegraphics[width=\linewidth]{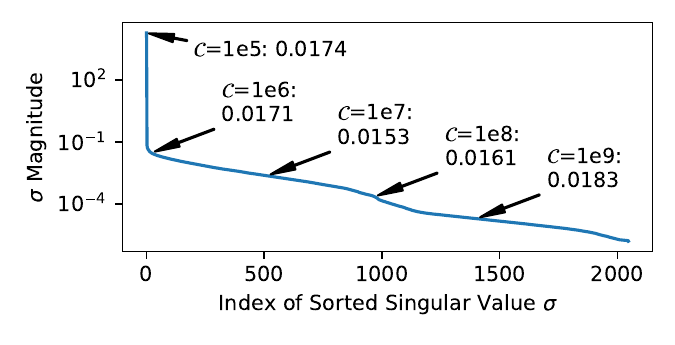}
  \caption{The curve indicates the magnitude of sorted singular values. We annotate five thresholds of the condition number, with the corresponding L1-norm differences of the solved model weights. }
  \label{fig:svd_values_plot}
\end{figure}
To deal with it, an upper threshold of the condition number is set as $\mathcal{C}$, meaning that singular values  $\sigma$ lower than  $\sigma_{max}/\mathcal{C}$ are truncated. Figure \ref{fig:svd_values_plot} shows the magnitude of the sorted singular values of the certain layer of Pythia-70m. Five thresholds (1e5, 1e6, 1e7, 1e8, and 1e9) and the corresponding L1-norm difference of the solved model weights compared to the original weight are annotated on the curve. As the threshold is gradually increased to retain more singular values, the L1-norm difference decreases. However, further increasing of the threshold causes the difference to grow instead, as numerical errors begin to counteract the benefits. We choose $\mathcal{C}$ as 1e7 throughout the experiments, which we find yields the most accurate solution in most cases.

\subsection{Perplexity of Extracted Model Weights}
To evaluate how the recovered weights perform on real tasks, we compare the perplexity of the extracted model weights with that of the victim models on the wikitext dataset~\cite{wikitext}.

Our results show that the extracted weights can already serve as an effective proxy model for downstream black-box attacks. After a short fine-tuning stage (at most 6 minutes), the recovered model closely matches the victim model’s perplexity and can effectively substitute it. Such near-lossless recovery substantially increases the attack severity, as it enables the attacker to obtain a model that behaves almost identically to the original and transform the following attacks into a white-box manner.

\begin{table}[tbp]
    \footnotesize
    \centering
    \caption{Perplexity on the Wikitext dataset. Lower perplexity indicates better performance.}
    \begin{tabular}{cccc}
    \toprule
    & \thead{Original \\ Model} & \thead{Stolen \\ Model} & \thead{Finetuned \\ Stolen Model} \\
    \midrule
    Pythia-70m & 31.81 & 44.46 & 32.43 \\
    GPT-2 & 21.11 & 47.92 & 21.15 \\
    \bottomrule
    \end{tabular}
\end{table}
\section{Related Work}
\label{sec:related_work}


\paragraph{Cryptography-based Secure Inference of Transformer Models}
Due to the rapidly growing concerns about data privacy in Transformer-based applications, numerous works have investigated two-party secure inference for the Transformer model.
Secure computation of nonlinear layers is significantly slower than that of linear layers~\cite{li2023mpcformer,THE-X,perm_dnn1,hao2022iron,zhang2024heprune}, and thus becomes the major performance bottleneck. This is because evaluating a nonlinear layer generally involves polynomial approximations or complex protocols, necessitating multiple communication rounds and extensive data transmission. The optimization of full-model encryption inference can be categorized into two main aspects as follows.

One type of work optimizes cryptographic protocols to improve the efficiency of secure inference.
CryptFlow2~\cite{cryptflow2} proposes a faster millionaire protocol to reduce the communication size of the comparison operation.
Some methods~\cite{rathee2021sirnn,gupta2023sigma,pang2023bolt} optimize secure lookup tables.
Another type of work modifies the Transformer model to tailor the cryptographic protocols.
Since accurate piecewise polynomials to approximate the nonlinear layers~\cite{dong2023puma,lu2023bumblebee} is costly, some works~\cite{chen2022x,zeng2023mpcvit,li2023mpcformer} use aggressive approximations of $\mathsf{Softmax}$ and $\mathsf{GELU}$.
However, such aggressive approximations lead to noticeable accuracy loss, even when employing knowledge distillation to mitigate the decline in accuracy.
Later works adopt fully homomorphic encryption and use GPU acceleration~\cite{zhang2025secure,park2025powerformer,moon2025thor}.
Despite these optimizations, nonlinear layers remain the main bottleneck.

\paragraph{Plaintext Nonlinear Layer Computation}
Due to the nonlinear layers being inherently unfriendly to cryptographic protocols, a workaround is to let the client evaluate the nonlinear layer in plaintext~\cite{GELU-Net,Bayhenn,attack_revealing,perm_dnn1,perm_dnn2,perm_transformer1}. 
These works allow the client to access intermediate activations for evaluating nonlinear layers. To prevent the exploitation of such activations, early approaches~\cite{GELU-Net,Bayhenn} restrict query access or employ Bayesian neural networks; however, they remain vulnerable to extraction attacks given sufficient queries~\cite{attack_revealing}. Subsequent works~\cite{perm_dnn1,perm_dnn2,perm_transformer1,perm_transformer3} observe that nonlinear computations are invariant to the positions of values in the activation tensor, and thus propose revealing shuffled activations to the client. Although shuffling disrupts activation structure and is widely considered secure in practice, this work demonstrates an attack that can still recover model weights (up to symmetries).

\section{Conclusion}
\label{sec:conclusion}
Shuffling defense in the \newpi{} inference demonstrates great potential for balancing security and performance within the Transformer model.
This work revisits the shuffling defense by introducing an attack that aligns shuffled activations and subsequently extracts model weights.
Through both theoretical analysis and experimental results, we demonstrate that our method can successfully extract model weights, with the L1-norm difference from the original weights ranging from $10^{-4}$ to $10^{-2}$.
By addressing these issues early, we aim to understand the limitations of the shuffling defense and contribute to its improvement, ultimately enhancing its security and reliability.


\section*{Limitations}
While our approach theoretically guarantees the feasibility of weight extraction and has been successfully validated on small-scale models, it remains subject to the inherent limitations of neural network-based attacks with respect to model scale. Specifically, as the target model size increases, the fidelity of the extracted weights tends to degrade—resulting in reduced recovery accuracy. 
However, it is important to emphasize that the goal of this work is not to facilitate malicious model extraction, but rather to expose previously overlooked vulnerabilities in existing defense mechanisms. From this perspective, successful validation on small-scale models is sufficient to fulfill our objective.

\section*{Acknowledgements}
This work was supported by the Shanghai Committee of Science and Technology, China (Grant No. 24BC3200900), the Shanghai Academy of Future Internet Technology, the National Natural Science Foundation of China (NSFC) (Grant No. 62532006), and the Shanghai Qi Zhi Institute Innovation Program (SQZ202316).

\bibliography{custom}

\newpage
\appendix
\section{Appendix}
\renewcommand\thesection{\Alph{section}}

\appendix

\section{Complete Attack Algorithm}
\label{appendix:complete_attack}
\begin{algorithm}[tbp]
    \caption{Weight Extraction Attack}
    \begin{algorithmic}[1]
        \renewcommand{\algorithmicrequire}{\textbf{Input:}}
        \REQUIRE Sequence $S$;

        \COMMENT{Prepare aligned dataset}
        \STATE Initialize $n$ as an appropriate value greater than hidden dimension $h$;
        \FOR{$i=1$ to $n$}
            \STATE $\mathsf{sf}(\mathbf{x}^{(1)}_i, \pi_1^{(i)}), \mathsf{sf}(\mathbf{x}^{(2)}_i, \pi_2^{(i)}), \cdots, \mathbf{y}_i^{(L)}=\mathcal{O}_{\pi}(S)$;
        \ENDFOR
        \FOR{$l=1$ to $L-1$}
            \STATE Align the dataset $\{\mathsf{sf}(\mathbf{x}^{(l)}_i, \pi_i^{(l)}) \}$ using  $\pi_{k}^{(l)}$ as the reference permutation and obtain $\{\mathsf{sf}(\mathbf{x}^{(l)}_i, \pi_k^{(l)})\}$;
            \STATE $\mathbf{X}^{(l)}\pi^{(l)}_k=concate(\{\{\mathsf{sf}(\mathbf{x}^{(l)}_i, \pi_k^{(l)}) \})$;
            \STATE Compute the pseudo-inverse $\pi^{{(l)}^{-}}_k\mathbf{X}^{{(l)}^{-}}$;
            \STATE $\mathbf{W}^{{(l)}^{\prime}} = \pi^{(l)^{-}}_k \mathbf{X}^{(l)^{-}} \mathbf{X}^{(l+1)} \pi^{(l+1)}_k= \pi^{(l)^{-}}_k \mathbf{W}^{(l)} \pi^{(l+1)}_k$, where $\mathbf{W}^{(l)}$ is the original weights;
            
        \ENDFOR

        \renewcommand{\algorithmicensure}{\textbf{Output:}}
        \ENSURE  Weight set $\{ \mathbf{W}^{{(l)}^{\prime}} \}$ of all linear layers;

\end{algorithmic} 
\label{alg:attack}
\end{algorithm}

With the insight of breaking shuffling defense and constructing proximate inputs through random truncation error, we formally describe the attack algorithm that can extract the model weights of the Transformer model.

The adversary first decides the number of queries $n$. Note that the query number should be at least the greatest size of the model, i.e., $4*d_{model}$, which is the dimension of the FFN module. This is because solving the linear system requires computing the inverse of the aligned input vectors. Thus, $n$ should be greater than $4*d_{model}$ for accurate matrix inverse (line 1).

Then, the adversary keeps feeding the model with the same prompt $S$. Due to the secure truncation, in operations such as the $\mathsf{LayerNorm}$ in the embedding layer, a small perturbation is introduced to the activation vectors, which the adversary expects to construct close activations.
In this way, the client collects a dataset that includes the shuffled intermediate activations of all linear layers $\{\mathsf{sf}(\mathbf{x}^{(l)}_i, \pi_i^{(l)}) \}$, for $i \in [n]$ and $l \in [L]$ (lines 2-4). Note that the model may generate many tokens in a response. We only collect tokens at the same generation step.

With the dataset, the adversary can align them using the method in Section 4. For each layer, the adversary arbitrarily selects the permutation scheme of a vector as the reference, which we refer to as $\pi^{(l)}_k$, and then aligns the other vectors according to $\pi^{(l)}_k$.
The aligned vectors are then concatenated as a matrix. The model weights can be computed using the inverse of the input matrix (lines 5-10).
Although the obtained weights $\mathbf{W}^{{(l)}^{\prime}}$ differ from the original weights $\mathbf{W}^{(l)}$ up to a row and column shuffling and this shuffling is unknown to the adversary, the solved weights still perform the original functionality as the consecutive layer adopts the same shuffling scheme. 

\section{Proof of the Equivalence between Equation (\ref{equ:transforming_pi}) and Equation (\ref{equ:matching})} 
\label{appendix:equivalence_proof}
This section demonstrates the equivalence between Equation (\ref{equ:transforming_pi}) and Equation (\ref{equ:matching}). We first recall some notations.  $\mathbf{x}_a^{\prime}, \mathbf{x}_b^{\prime} \in \mathbb{R}^h$ are two differently shuffled but value-close activation vectors. $\mathbf{M} \in \{0,1\}^{h \times h}$ is a permutation matrix that has only single one in each row or column.

We begin from the Equation (\ref{equ:transforming_pi})
\begin{equation}
\resizebox{\columnwidth}{!}{$
\left\| \mathbf{x}_a^{\prime} -  \mathbf{x}_b^{\prime} \mathbf{M} \right\|_2^2 = \sum_{i=1}^h \left( \mathbf{x}_a^{\prime}[i] - \sum_{j=1}^h \mathbf{x}_b^{\prime}[j] \mathbf{M}[j,i] \right)^2.
$}
    \label{equ:derive_transform_pi}
\end{equation}

Since $\mathbf{M}$ only has single one in each column, we have $\sum_{j=1}^h \mathbf{x}_b^{\prime}[j] \mathbf{M}[j,i] = \mathbf{x}_b^{\prime}[\sigma(i)]$, where $\sigma(i)$ is an index function indicates the row index of 1 in $i_{th}$ column of $\mathbf{M}$. Then Equation \eqref{equ:derive_transform_pi} becomes
\begin{equation}
\resizebox{\columnwidth}{!}{$
    \begin{aligned}
        &  \sum_{i=1}^h \left( \mathbf{x}_a^{\prime}[i] - \mathbf{x}_b^{\prime}[\sigma(i)] \right)^2\\
        = & \sum_{i=1}^h \left[ \left( \mathbf{x}_a^{\prime}[i] - \mathbf{x}_b^{\prime}[\sigma(i)] \right)^2 + 0*\sum_{j=1,j\neq i}^h \left( \mathbf{x}_a^{\prime}[j] - \mathbf{x}_b^{\prime}[\sigma(j)] \right)^2 \right] \\
        = & \sum_{i,j\in \mathcal{H}} \mathbf{M}[i,j] \cdot (\mathbf{x}_a^{\prime}[i] - \mathbf{x}_b^{\prime}[j])^2.
    \end{aligned}
    $}
\end{equation}
This is exact the objective $\sum_{i,j\in \mathcal{H}}\mathbf{M}[i, j] \odot \mathbf{D}[i, j]$ in Equation (\ref{equ:matching}). \hfill $\square$

\section{Equivalent Functionality of Shuffled Weights}
\label{appendix:proof_equivalent_weights}
As stated in Section \ref{subsec:expose_weights}, for shuffled weights of consecutive layers, functional equivalence with the original weights is preserved if the output dimension of the preceding layer and the input dimension of the subsequent layer adhere to the same permutation. We use two consecutive layers to illustrate this and the same idea can be extended to the entire model.

Let $\mathbf{W}^{(l)}$ and $\mathbf{W}^{(l+1)}$ denote the weights of two consecutive layers. For a given input vector $\mathbf{x}^{(l)}$ to layer $l$, the forward propagation is expressed as
\begin{equation}
    \mathbf{x}^{(l+2)}=g(\mathbf{x}^{(l)}\mathbf{W}^{(l)})\mathbf{W}^{(l+1)}
\end{equation}
The $g(\cdot)$ are nonlinear layers, such as the $\mathsf{GELU}$, $\mathsf{Softmax}$, or $\mathsf{Layernorm}$.

In the proposed attack, the solved weights differ from the original weights up to a row-wise and column-wise shuffling, which are $\mathbf{W}^{{(l)}^{\prime}} = \pi^{(l)^{-}} \mathbf{W}^{(l)} \pi^{(l+1)}$ and $\mathbf{W}^{{(l+1)}^{\prime}} = \pi^{(l+1)^{-}} \mathbf{W}^{(l+1)} \pi^{(l+2)}$. The input to $\mathbf{W}^{(l)}$ is $\mathbf{x}^{(l)}\pi^{(l)}$ since the output channel of prior weights $\mathbf{W}^{{(l-1)}^{\prime}} = \pi^{(l-1)^{-}} \mathbf{W}^{(l-1)} \pi^{(l)}$ is also shuffled. Then the forward propagation using the equivalent weights is 
\begin{equation}
\resizebox{\columnwidth}{!}{$
\begin{aligned}
    & g(\mathbf{x}^{(l)}\pi^{(l)}\mathbf{W}^{{(l)}^{\prime}})\mathbf{W}^{{(l+1)}^{\prime}}\\
    = & g(\mathbf{x}^{(l)}\pi^{(l)}\pi^{(l)^{-}} \mathbf{W}^{(l)} \pi^{(l+1)})\pi^{(l+1)^{-}} \mathbf{W}^{(l+1)} \pi^{(l+2)} \\
    = & g(\mathbf{x}^{(l)} \mathbf{W}^{(l)} )\pi^{(l+1)}\pi^{(l+1)^{-}} \mathbf{W}^{(l+1)} \pi^{(l+2)} \\
    = & g(\mathbf{x}^{(l)} \mathbf{W}^{(l)} ) \mathbf{W}^{(l+1)} \pi^{(l+2)} \\
    = & \mathbf{x}^{(l+2)} \pi^{(l+2)} 
\end{aligned}
$}
\end{equation}
The equality in the third line is because the permutation invariance property of the activation function $g(\cdot)$, allowing the permutation matrix $\pi^{(l+1)}$ to be factored out. Subsequently, the permuted output $\mathbf{x}^{(l+2)} \pi^{(l+2)}$ serves as input to the next shuffled weight matrix $\mathbf{W}^{{(l+2)}^{\prime}} = \pi^{(l+2)^{-1}} \mathbf{W}^{(l+2)} \pi^{(l+3)}$. This formulation ensures that the same permutation between the output dimension of $\mathbf{W}^{(l)}$ and the input dimension of $\mathbf{W}^{(l+1)}$ preserves the functional equivalence of the shuffled weight matrices.

\section{Dealing with the Multi-head Attention}
\label{subsec:handle_multi_head_attention}
Recovering weights becomes more complicated when the matrix multiplication of activations in the attention module is privately computed in some studies~\cite{perm_transformer1}.
In such a case, although the adversary cannot directly obtain the input and output of the linear layer, the activation matrix multiplication still preserves the linear system, allowing for the derivation of equivalent weights that function identically to the original linear layers.
However, the multi-head attention mechanism increases the complexities of the linear system, as detailed below. 
For ease of discussion, we first define additional notations.
The weight of layer $Linear_{qkv}$ is decomposed into weights $\mathbf{W}_q$, $\mathbf{W}_k$, and $\mathbf{W}_v$. For the token being generated, the input and output of $\mathbf{W}_q$ are $\mathbf{x}$ and $\mathbf{q}$. For prompt and already generated tokens, the input matrix of $\mathbf{W}_k$ and $\mathbf{W}_v$ are denoted by $\mathbf{X}_{pre}$ with output $\mathbf{K}$ and $\mathbf{V}$, which are known as the KV cache. 

\paragraph{Equivalent Weights of $\mathbf{W}_{q}$ and $\mathbf{W}_{k}$}
For the $h_{th}$ attention head, we define the equivalent weights $\mathbf{W}_{qk}^h = \mathbf{W}_q^h  {\mathbf{W}_k^h}^T$. Different heads represent the independent sub-problems that can be solved in the same way. The adversary holds the input $\mathbf{x}$ to $\mathbf{W}_{q}$, the prefix input $\mathbf{X}_{pre}$ to $\mathbf{W}_{k}$, and the input $\mathbf{s}$ of the $\mathsf{softmax}$ (override the notation meaning model input in prior sections). According to Equation \eqref{equ:attention}, their relationship is formulated as $\mathbf{s}^h=\mathbf{x}  \mathbf{W}_q^h  {\mathbf{W}_k^h}^T  \mathbf{X}_{pre}^T$.
The problem is still a linear problem through the transformation: 
\begin{equation}
    \mathbf{s}^h=\mathbf{x}  \mathbf{W}_{qk}^h  \mathbf{X}_{pre}^T=\left(\mathbf{X}_{pre} \otimes \mathbf{x}\right)^T \operatorname{vec}(\mathbf{W}_{qk}^h),
    \label{equ:qk}
\end{equation}
where the $\otimes$ is the Kronecker product and $\operatorname{vec}(\cdot)$ means vectorizing the matrix. 
The adversary can first compute the Kronecker product, resulting in a matrix with a hidden dimension equal to the square of the model's hidden dimension.
Then, it becomes the same linear system as Equation \eqref{equ:general_linear_system} to solve the equivalent weight $\mathbf{W}_{qk}^h$. 

\paragraph{Equivalent Weights of $\mathbf{W}_{v}$ and $\mathbf{W}_{o}$}
The equivalent weight for $\mathbf{W}_{v}$ and $\mathbf{W}_{o}$ is given by $\mathbf{W}_{vo} = \left[\begin{array}{c}\mathbf{W}_v^1 \mathbf{W}_o^1 \\ \mathbf{W}_v^2 \mathbf{W}_o^2 \\ \cdots \\ \mathbf{W}_v^H \mathbf{W}_o^H\end{array}\right]$, where the superscripts indicate the head index. In this case, the adversary holds the prefix input $\mathbf{X}_{pre}$ to $\mathbf{W}_{v}$, the softmax output $\mathbf{p}$, and the $\mathbf{W}_{o}$ output $\mathbf{o}$. 
Due to the multi-head mechanism, the original relationship $\mathbf{o}=\mathbf{p} \cdot \mathbf{X}_{pre} \mathbf{W}_v \cdot \mathbf{W}_o$ in Equation \eqref{equ:attention} transforms into:
\begin{equation}
  \resizebox{\columnwidth}{!}{$
    \begin{aligned}
    \mathbf{o} & =\text { Concat }\left(\mathbf{head}_1, \ldots, \mathbf{head}_H\right) * \mathbf{W}_o \\
    & =\left[\mathbf{p}^1 \mathbf{X}_{pre} \mathbf{W}_v^1 \mid \mathbf{p}^2 \mathbf{X}_{pre} \mathbf{W}_v^2 \mid \ldots \mid \mathbf{p}^H \mathbf{X}_{pre} \mathbf{W}_v^H\right] * \mathbf{W}_o \\
    & =\left[\mathbf{p}^1 \mathbf{X}_{pre} \mid \mathbf{p}^2 \mathbf{X}_{pre} \mid \ldots \mid \mathbf{p}^H \mathbf{X}_{pre}\right] *\left[\begin{array}{c}
    \mathbf{W}_v^1 \mathbf{W}_o^1 \\
    \mathbf{W}_v^2 \mathbf{W}_o^2 \\
    \cdots \\
    \mathbf{W}_v^H \mathbf{W}_o^H
    \end{array}\right]
    \end{aligned}
    $}
    \label{equ:vo}
\end{equation}
Note that the $\left[\mathbf{p}^1 \mathbf{X}_{pre} \mid \ldots  \mid \mathbf{p}^H \mathbf{X}_{pre}\right]$ is a vector whose dimension is $H$ times the model dimension. Finally, the equivalent weight $\mathbf{W}_{vo}$ is solved in the same way as Section \ref{subsec:expose_weights}.

\begin{figure*}[tbp] 
	\centering
    \includegraphics[width=0.48\linewidth]{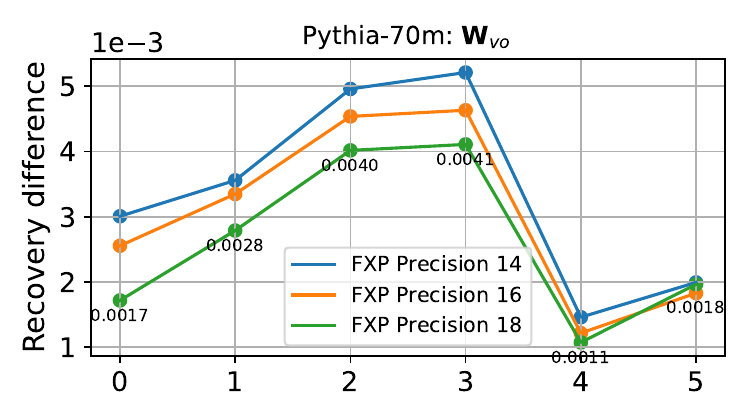}
    \includegraphics[width=0.48\linewidth]{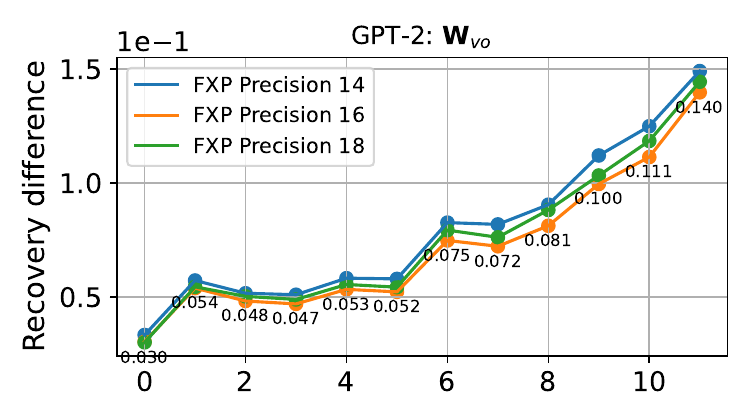}

	\caption{The attacking results on the equivalent weights $\mathbf{W}_{vo}$.
	}
	\label{fig:e2e_attn_attack_results}
\end{figure*}
\paragraph{Evaluation}
Figure \ref{fig:e2e_attn_attack_results} shows the results when the activation matrix multiplications are privately computed. Results are presented only for the equivalent weight $\mathbf{W}_{vo}$.
The L1-norm differences for Pythia-70m are smaller than 0.004 and the differences for GPT-2 are smaller than 0.14. The larger differences in GPT-2 are due to its relatively higher number of attention heads (12), making the linear system more challenging to solve.
Recovering equivalent weights $\mathbf{W}_{qk}$ remains a challenge. Computing the Kronecker product makes the input dimension in Equation \eqref{equ:qk} become the square of the model dimension, such as $768^2$. We leave this for future work.

\section{Possible Defenses}
\label{sec:defenses}
Our attack identifies the weaknesses of the shuffling defense in the existing \newpi{} inference. 
We also briefly list corresponding defense that can mitigate the proposed attack. 

\paragraph{Noise Addition}
The first defense strategy is to increase the alignment error by adding noise. Adding noise is a common technique used in neural network defense to obscure leaked information. However, noise addition is generally constrained within a threshold to prevent degradation of model accuracy, meaning this approach can only partially mitigate the attack.

\paragraph{Partial Plaintext Nonlinear Layer Computation}
The second strategy is to make only part of the nonlinear layers publicly computed. We observe that the primary computational cost mainly arises from the $\mathsf{Softmax}$ and $\mathsf{GELU}$ layers, while the $\mathsf{Layernorm}$ is much faster. This is because the nonlinear operation, the reciprocal square root, of the $\mathsf{Layernorm}$ only computes once per vector~\cite{lu2023bumblebee,pang2023bolt}.
The defender can tolerate a slight increase in latency to make the $\mathsf{Layernorm}$ is computed privately. This approach ensures that for all types of linear layers in the Transformer decoder, either the input or output remains unknown to the adversary, making weight recovery impossible.

\end{document}